\documentclass[
 aip,
 amsmath,amssymb,
 reprint,
]{revtex4-1}

\usepackage{graphicx,amsmath,bm,color}
\usepackage[utf8]{inputenc}
\usepackage[T1]{fontenc}
\usepackage{mathptmx}
\usepackage{etoolbox}
\usepackage[version=4]{mhchem}
\usepackage{color}
\usepackage[normalem]{ulem}
\usepackage{url}
\usepackage{algorithm}
\usepackage{algpseudocode}

\makeatletter
\def\@email#1#2{%
 \endgroup
 \patchcmd{\titleblock@produce}
  {\frontmatter@RRAPformat}
  {\frontmatter@RRAPformat{\produce@RRAP{*#1\href{mailto:#2}{#2}}}\frontmatter@RRAPformat}
  {}{}
}%
\makeatother
\algnewcommand{\LineComment}[1]{\State \(\triangleright\) #1}

\begin{document}

\preprint{AIP/123-QED}

\title[Prevalence of multistability and nonstationarity in driven chemical networks]{Prevalence of multistability and nonstationarity in driven chemical networks}

\author{Zachary G. Nicolaou}
\affiliation{Department of Chemistry, University of Massachusetts Boston, Boston, Massachusetts 02125, USA}
\affiliation{Department of Applied Mathematics, University of Washington, Seattle, WA 98195, USA}

\author{Schuyler B. Nicholson}
\affiliation{Department of Chemistry, University of Massachusetts Boston, Boston, Massachusetts 02125, USA}
\affiliation{Department of Physics, University of Massachusetts Boston, Boston, Massachusetts 02125, USA}
\affiliation{Department of Chemistry, Northwestern University, Evanston, Illinois 60208, USA}

\author{Adilson E. Motter}
\affiliation{Department of Physics and Astronomy, Northwestern University, Evanston, Illinois 60208, USA}
\affiliation{Northwestern Institute on Complex Systems, Northwestern University, Evanston, Illinois 60208, USA}
\affiliation{Department of Engineering Sciences and Applied Mathematics, Northwestern University, Evanston, Illinois 60208, USA}

\author{Jason R. Green}
\affiliation{Department of Chemistry, University of Massachusetts Boston, Boston, Massachusetts 02125, USA}
\affiliation{Department of Physics, University of Massachusetts Boston, Boston, Massachusetts 02125, USA}
\email{jason.green@umb.edu}

\date{\today}

\begin{abstract}

External flows of energy, entropy, and matter can cause sudden transitions in the stability of biological and industrial systems, fundamentally altering their dynamical function. How might we control and design these transitions in chemical reaction networks?
Here, we analyze  transitions giving rise to complex behavior in random reaction networks subject to external driving forces. In the absence of driving, we characterize the uniqueness of the steady state and identify the percolation of a giant connected component in these networks as the number of reactions increases.
When subject to chemical driving (influx and outflux of chemical species), the steady state can undergo bifurcations, leading to multistability or oscillatory dynamics.
By quantifying the prevalence of these bifurcations, we show how chemical driving and network sparsity tend to promote the emergence of these complex dynamics and increased rates of entropy production.
We show that catalysis also plays an important role in the emergence of complexity, strongly correlating with the prevalence of bifurcations. Our results suggest that coupling a minimal number of chemical signatures with external driving can lead to features present in biochemical processes and abiogenesis.\\[1.5em]

\noindent J. Chem. Phys. 158, 225101 (2023)\\
\noindent DOI: \url{https://doi.org/10.1063/5.0142589}

\end{abstract}

\maketitle

\section{Introduction}
\label{introsec}
Characterizing the emergence of complex dynamical behaviors in networks of chemical reactions is critical to our understanding of natural and industrial processes, from abiogenesis~\cite{1952_Turing} and morphogenesis~\cite{2001_Dittrich} to combustion~\cite{2010_Law,2017_Newcomb,2018_Newcomb} and atmospheric pollution~\cite{2008_Atkinson}. For example, the dynamical mechanisms of oscillatory reactions, such as the Belousov-Zhabotinsky reaction~\cite{1993_Cross}, and chemically active matter~\cite{2015_Boekhoven} can generate universal and functional patterns. In both synthetic and biological systems, the appearance of multistability and/or nonstationarity~\cite{2020_Nicolaoub} in the chemical kinetics underlies functions such as memory, evolution, and responsiveness to environmental conditions. However, an outstanding problem is how external conditions trigger the nontrivial chemical dynamics in reaction networks. While some understanding has been distilled from minimal models, it is unclear how prevalent these dynamics are or how they might be designed.

Random matrix theory has been one significant approach for assessing the stability and universality of complex networks. Early work motivated by ecological analogs of chemical networks such as the Lotka-Volterra model~\cite{1971_Goel}, suggested that complexity in the network structure would inhibit the stability of dynamical states~\cite{1972_May}. These early developments relied on an (over)simplified theory of dense matrices with uncorrelated matrix elements. Only later was it recognized that the correlations of interactions and other non-random factors can in fact favor a variety of phenomena leveraged by living systems for biological function~\cite{2016_Amir,2016_Grilli}. While significant progress has been achieved in sparse random matrix theory with uncorrelated elements~\cite{2019_Metz}, the effects of correlations encoded by Laplacian interactions, for example, continue to attract attention~\cite{Neri_2016,2018_Forrow}. Of specific interest here, is how externally imposed driving forces couple to correlations following from the structures of chemical reaction networks.

Among the features of chemical networks that lead to complexity,  catalysis (in which a chemical species acts to accelerate a reaction rate by acting as both a reactant and a product) has long been thought to play a central role. This hypothesis is bolstered by the prominence of catalysis in classical models of complexity, such as the Brusselator and Oregonator models for the Belousov-Zhabotinsky chemical reaction~\cite{1974_Field,2015_Showalter}. Catalysis also plays a central role in living systems, motivating efforts to understand its effects on artificial biologically inspired reaction networks~\cite{2010_Riehl,2017_himeoka,2019_Kroiss,2019_Epstein,2021_Mondal}. While recent progress has been made in characterizing catalysis in arbitrary networks~\cite{2020_Blokhuis}, the general problem of defining the extent of catalysis in a network remains open. Specific forms of driving also play a role in the development of complexity by producing multistability and persistent oscillatory states in random chemical networks~\cite{2017_Horowitz}, in contrast to the unique steady state observed in thermodynamic equilibrium. When coupled with diffusion, local concentration inhomogeneities can be amplified through (auto)catalysis if a small increase of one species over its homogeneous steady state concentration induces a further increase in concentration. Catalysis can also be indirect where the catalyst accelerates the production of a product through intermediate reactions with other substances.

Here, we computationally drive the formation of complexity in the form of multistability and nonstationarity in artificial chemical networks, analyzing both its emergence and prevalence. Specifically, we consider the impacts of the number of species and reactions, the proportion of catalytic reactions, and the strength of chemical driving on the emergence of complex chemical dynamics in random chemical reaction networks.
In Sec. ~\ref{networksec}, we define a physically relevant class of chemical networks, and we define the soft chemostat that we use to drive these networks in Sec.~\ref{thermosec}.  We relate our networks to graphs in Sec.~\ref{graphsec}, characterize the steady states of these networks in the absence of driving in Sec.~\ref{steadysec}, and note the important role of atomic conservation laws in Sec.~\ref{conservationsec}. We then describe percolation transitions that occur as the number of reactions increases in Sec.~\ref{percolationsec}. 
In Sec.~\ref{complexitysec}, we detail the transitions to complex behavior through bifurcations caused by external driving.
In Sec. \ref{factorssec}, we analyze the role of the total number of reactions and the proportion of catalytic reactions on these transitions, uncovering a thermodynamic tendency to increase the rate of entropy production during bifurcations.
Finally, we discuss the implications of the results in Sec.~\ref{discussionsec}.

\section{Results}
\subsection{Reaction networks}
\label{networksec}
Our model chemical reaction networks consist of a set of possible reactions between a set of chemical species $\{\ce{A_i}\}$ that can occur within a reaction vessel, as shown in Fig.~\ref{fig1}(a).
We generate these networks randomly (as detailed below and summarized in Appendix \ref{adetailed}), choosing $n$ chemical species and $\lfloor rn \rfloor$ chemical reactions of the form,
\begin{equation}
\label{eq}
  \sum_i \nu_{i \ell} \ce{A_i <-->[\kappa^+_\ell][\kappa^-_\ell]} \sum_i\eta_{i \ell} \ce{A_i}.
\end{equation}
The parameter $r$ controls the number of reactions in the network, which has important consequences for the dynamics. These reactions are defined by stoichiometric reactant and product vectors $\nu_{i \ell}$ and $\eta_{i \ell}$ and the forward and reverse reaction rate coefficients $ \kappa^+_\ell $ and $ \kappa^-_\ell $, where $\ell$ is the reaction index and $i$ is the species index (throughout, for notational brevity, we do not distinguish between vectors/matrices and their components). The reaction vessel is at a constant temperature and pressure so that the reaction rate coefficients are time-independent. We assume the rate of the forward and reverse reactions in Eq.~\eqref{eq} is determined by mass action kinetics as $j^+_\ell \equiv  \kappa^+_\ell \prod_{i} X_i^{\nu_{i \ell}}$ and $j^-_\ell \equiv  \kappa^-_\ell \prod_{i} X_i^{\eta_{i \ell}}$, respectively, where $X_i$ denotes the concentration of $\ce{A_i}$.

\begin{figure*}[t]
\begin{center}
\includegraphics[width=\linewidth]{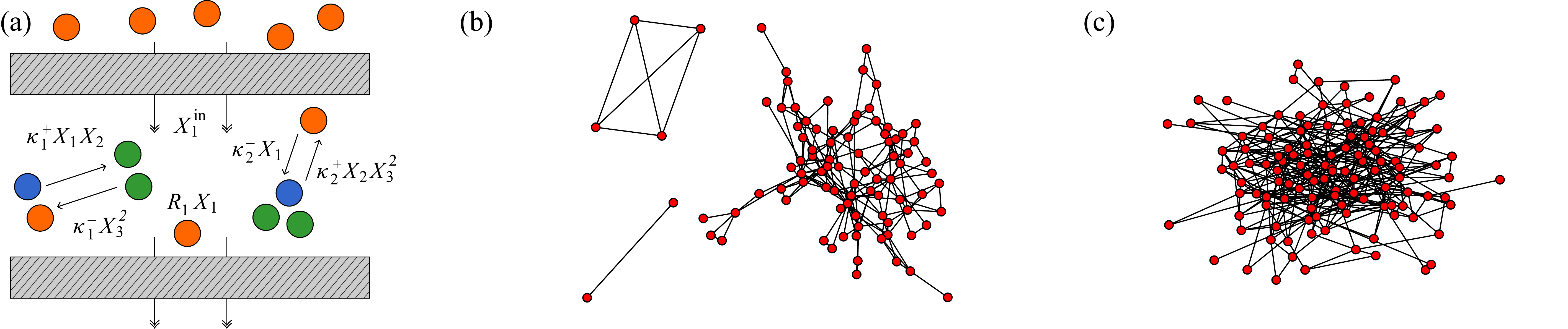}
\end{center}
\caption{Random reaction networks. (a) Chemical network, defined as a collection of random reversible chemical reactions. Reaction rates, governed by mass-action kinetics, are chemically driven through mass influxes and removal rates. {(b) and (c) Projection graph of sample networks with nodes representing chemical species and links representing reactions. When the number of reactions is small (b), the network is decomposed into multiple connected components. For a sufficiently large number of reactions (c), all nodes  are in a single  connected component. Networks in (b) and (c) are generated with $n=128$ chemical species, $r=0.48$ in (b) and $r=0.97$ in (c), and isolated nodes representing species that do not participate in any reactions are not depicted.} \label{fig1}}
\end{figure*}

\subsection{{Chemical driving}}
\label{thermosec}
A soft chemostat acts on these networks through driving terms that push selected species toward prescribed concentrations. We model this chemostat through source terms (which add species into the system at fixed rates) and sink terms (which extract them at a rate dependent on their concentrations), as indicated in Fig.~\ref{fig1}(a). 

Crucially, this driving differs from an open exchange with a single chemical reservoir at a prescribed chemical potential.  In particular,  the grand canonical potential describes the irreversible thermodynamic relaxation to a unique equilibrium under open exchange with a single reservoir at fixed chemical potential, while our chemical networks may relax to nonunique steady states or to persistent oscillatory or chaotic dynamics.   Our driving can be described as an open exchange with two or more chemical reservoirs with incompatible chemical potentials, much like a heat engine operating between thermal reservoirs at differing temperatures.  The differing chemical potentials and rate constants governing the exchanges permit the influx and outflux rates to violate local detailed balance, leading to a perpetual state of nonequilibrium.
The external drive is thus defined by independent influx and outflux rates $X_i^{\mathrm{in}}$ and $R_i X_i$. A number $\lfloor dn \rfloor$ of species in the network are randomly selected to be driven in this fashion, while $X_i^\mathrm{in}$ and $R_i$ are set to zero for the remaining species.

The rate of change of the concentrations is then given by the set of ordinary differential equations
\begin{align}
\label{kinetics}
\frac{{\mathrm d}X_i}{{\mathrm d}t} = X_i^{\mathrm{in}} - R_i X_i + \sum_\ell \left( \eta_{i \ell} - \nu_{i \ell} \right)\left(j^+_\ell - j^-_\ell\right).
\end{align}
We relate the driving parameters to a driving magnitude $\varepsilon$ and an initial target concentration $Z_i^0$ via $X_i^{\mathrm{in}}/R_i=(1+\varepsilon)Z_i^0$ (see Appendix \ref{aquasistatic}). The driving terms $X_i^{\mathrm{in}} - R_i X_i$ vanish as $X_i\to Z_i\equiv X^{\mathrm{in}}_i/R_i$, so the drive can be considered as an external force pushing the concentrations to the fixed level $Z_i$. 

\subsection{{Chemical reaction graphs}}
\label{graphsec}
In the theory of chemical reaction networks~\cite{2019_Feinberg}, the species-reaction graph encodes the structural features of the chemical reaction mechanism. This representation consists of a bipartite network with directed edges connecting species nodes and reaction nodes according to the species' membership as reactants and products. Equivalent higher-order representations, such as hypergraph or simplicial models, 
can also be employed\cite{2021_Battiston}, giving different perspectives on the structure of this network. The species-reaction graph contains all the structural information about a chemical reaction network, but simpler pairwise network projections have been defined in order to quantify percolation transitions for higher-order networks in other contexts~\cite{2009_Ghoshal,2009_Bradde,2021_Sun}.

Following these previous approaches to quantify percolation transitions,  we define a very simple projection of the chemical reaction network to an undirected graph consisting of nodes corresponding to chemical species (as shown in Fig.~\ref{fig1}). To do so, note that the rate of change of $X_i$ in Eq.~\eqref{kinetics} is determined by the rate of the reactions in which $\ce{A}_i$ is either a reactant or a product. The rates of these reactions are, in turn, determined by the concentrations $X_j$ of their respective reactants $\ce{A}_j$, and so it is natural to include a directed edge from $\mathrm{A}_j$ to $\mathrm{A}_i$ in the network. Since all reactions in the network are reversible, all such links would be bidrectional, so two species nodes can be connected with an undirected link if they participate in the same reaction in the network. The set of reactions thus defines an adjacency matrix for the network, which can be used to construct the graph of connected species, as shown in Figs.~\ref{fig1}(b)--\ref{fig1}(c). Although this adjacency matrix is random, its structure differs from traditional random matrices in that links between nodes are not introduced independently but in groups depending on the reaction order. For example, reversible bimolecular reactions of the form $\ce{A_i + A_j  <--> A_k + A_l}$ give rise to a large number of fully connected four-node subgraphs in the network regardless of the network size, whereas such subgraphs become increasingly infrequent in Erd\H{o}s-R\'{e}nyi graphs as the network size increase.

\subsection{{Undriven steady states}}
\label{steadysec}
We assume that each species can be assigned a standard Gibbs energy $G_i$, and we take a local detailed balance assumption that the ratio of the forward and reverse reaction rate constants (the equilibrium constant) is given by the change in Gibbs energy (see Appendix \ref{adetailed}).
In the absence of driving ($X_i^{\mathrm{in}}=R_i=0$), the detailed balance assumption implies that a steady state exists for each choice of conserved quantities (see Sec.  \ref{conservationsec}). In fact, it is easy to find one such steady state, $X_i^0\equiv \exp({-G_i/RT})$, which is determined by the Gibbs energies. The possibility of the nonuniqueness of this steady state solution has long been a central problem in the theory of chemical reaction networks~\cite{2006_craciun}. It is well known both in stochastic thermodynamics and linear irreversible thermodynamics that undriven chemical reaction networks exhibit dynamics with a Lyapunov function and a corresponding gradient form\cite{2010_Ge,2015_Anderson,2017_Mielke}, relaxing to a unique steady state for each stoichiometrically-compatible initial condition determined by the network conservation laws. For completeness, we derive this result for our networks in Appendix \ref{agradient}.

The stability of any steady state solution is determined by the spectrum of the Jacobian matrix, whose elements are given by:
\begin{align}
\label{jacobian}
J_{ij} &= -R_i \delta_{ij}+ \sum_\ell \kappa^+_\ell \left( \eta_{i \ell} - \nu_{i \ell} \right)\nu_{j \ell}X_j^{-1} \prod_{k} X_k^{\nu_{k \ell}} \nonumber \\
&+ \sum_\ell \kappa^-_\ell \left( \nu_{i \ell} - \eta_{i \ell} \right)\eta_{j \ell}X_j^{-1}\prod_{k} X_k^{\eta_{k \ell}}.
\end{align}
The elements of the Jacobian can be considered as an assignment of link weights on the adjacency network in Fig.~\ref{fig1}(b) for any given state.
Although the Jacobian  $J_{ij}^0$ at the equilibrium $X_i^0$ (see Appendix \ref{agradient}) is not a symmetric or even a normal matrix, we observe numerically that it has strictly real and nonpositive eigenvalues for every network that we generate~\cite{github}. In fact, given the gradient form of the linearized dynamics, the Jacobian for the undriven system is guaranteed to have strictly real, nonpositive eigenvalues. 
The symmetry in the network dynamics encoded by the gradient form is analogous to the $PT$ symmetry in dissipative quantum Hamiltonians. The $PT$ symmetry concerns invariance upon a simultaneous reversal in time and parity, which can guarantee real observables and unitary dynamics in systems that are not invariant under time reversal alone~\cite{2007_Bender}. For chemical reaction networks,  driving can break the analogous network symmetry, as we will show below, enabling the emergence of multistability and nonstationarity.

\subsection{{Conservation laws and random networks}}
\label{conservationsec}
Because the undriven dynamics can be expressed in a gradient form and the Jacobian is negative semidefinite  (see Appendix \ref{agradient}), the undriven steady state is unique so long as zero is not an eigenvalue of the Jacobian. Furthermore, zero is an eigenvalue of the undriven Jacobian if and only if the span of the $n_r$ stoichiometric vectors $\eta_{i\ell}-\nu_{i\ell}$ has dimension less than the number of chemical species $n$. Thus, the problem of determining if undriven the steady state is unique reduces to the purely structural problem of determining the rank of the stoichiometric matrix $\eta_{i\ell}-\nu_{i\ell}$. By the argument leading to Eq.~\eqref{negative} with $k_{\ell}^+=1$ and $G_k=0$, the rank deficiency of $\eta_{i\ell}-\nu_{i\ell}$ is equivalent to the singularity of the square matrix $M_{ij} = \sum_{\ell} (\eta_{i\ell}-\nu_{i\ell})(\nu_{j\ell}-\eta_{j\ell})$. When the model possesses conserved quantities, we expect a corresponding non-uniqueness in the steady state. 

In the present context, a conservation law corresponds to a set of weights $l_i$ such that $\sum_i(\eta_{i\ell}-\nu_{i\ell})l_i=0$, because $L\equiv\sum_i l_iX_i$ is then a conserved quantity\cite{2016_rao,2018_Rao,2022_Cengio}. Conservation laws emerge generically in chemical reaction networks because of stoichiometric constraints such as atomic conservation or the presence of conserved moieties, and to assess stability, we restrict attention to chemically accessible perturbations that do not alter the conserved quantities. Since we often expect a conservation law for each atomic element in the model,  we equate the number of imposed conservation laws with the number of atomic elements in the model.  We, therefore, expect the rank deficiency of $M_{ij}$ to be $n-a$, where $a$ is the number of atomic elements in the model.  Additional conservation can arise, both because the rank of the stoichiometric matrix can be smaller than $n-a$ for sufficiently sparse networks and because of the presence of physically significant cycles in the reaction network.
\begin{figure*}[t]
\begin{center}
\includegraphics[width=0.9\linewidth]{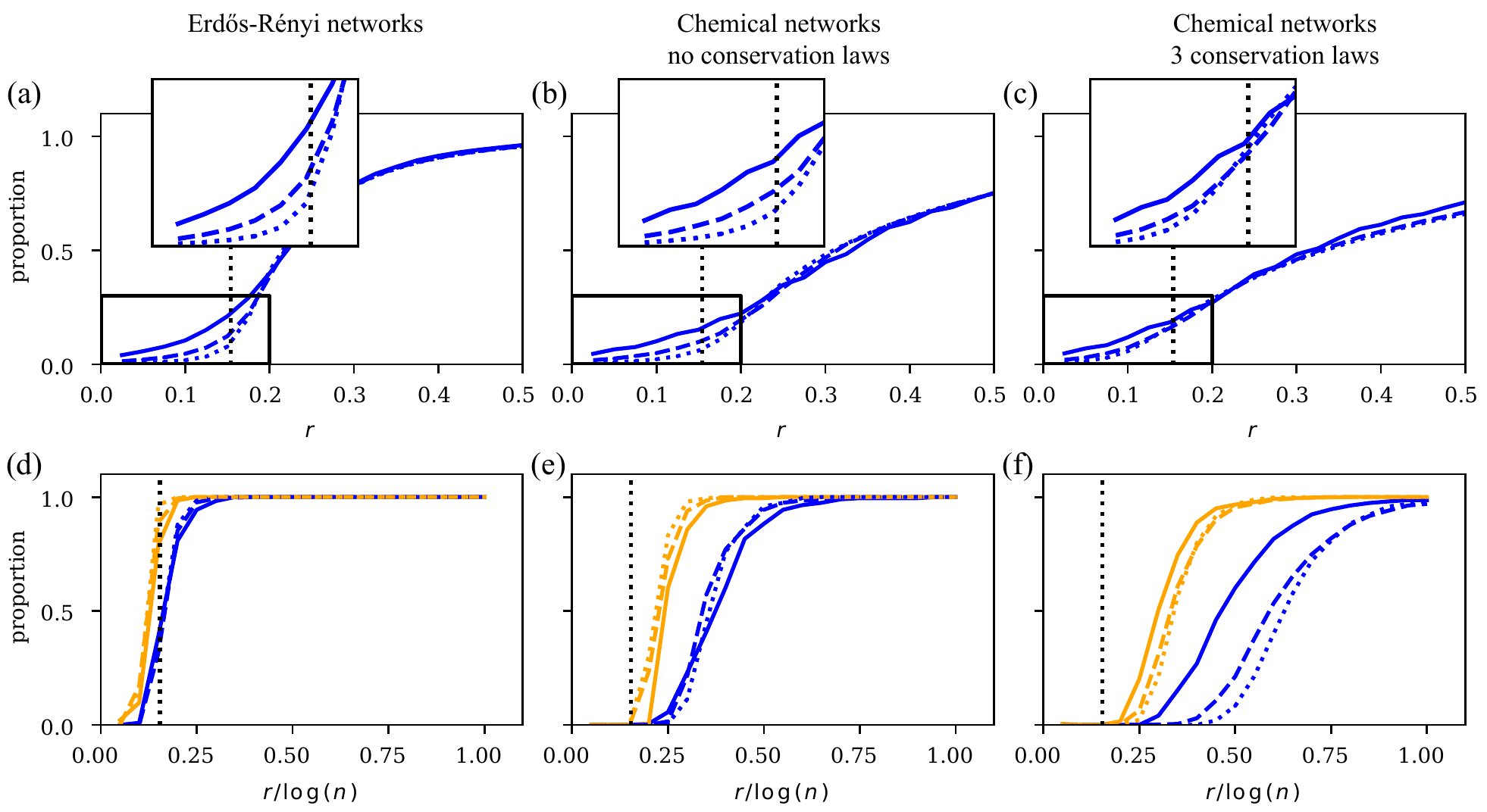}
\end{center}
\caption{Percolation transitions in Erd\H{o}s-R\'enyi and chemical networks. (a)-(c) Average proportion of the network in the largest connected component for Erd\H{o}s-R\'enyi networks (a) and chemical networks with no conservation laws (b) and $a=3$ conservation laws (c), where the insets show a blow-up of the percolation of the giant connected components. (d)-(f) The proportion of networks that are fully connected (blue lines) and networks with full-rank $M_{ij}$ restricted to the giant connected component (orange lines) for Erd\H{o}s-R\'enyi networks (d) and chemical networks with $a=0$ (e) and $a=3$ (f).  In all cases, plots show the results for $1024$ network realizations with $n=64$ (solid colored lines) $n=256$ (dashed colored lines) and $n=1024$ (dotted colored lines), and the dotted black lines mark the predicted transition threshold for Erd\H{o}s-R\'enyi networks. \label{fig2}}
\end{figure*}

{Our algorithm for generating a random chemical network with a prescribed number of species $n$, reactions $\lfloor rn \rfloor$, and atomic conservation laws $a$ is presented in detail in Appendix \ref{adetailed}. For each molecular species, we first randomly assign an atomic count $\ce{A}_{ik}$ for the number of atoms of type $k$ in the chemical formula for $\ce{A}_i$. The number of atoms and the distribution of their appearance among the molecular species are features of the reaction network that will vary depending on the application area.  For simplicity, throughout we restrict to $a=3$ conservation laws, with $\ce{A}_{ik}$ randomly sampled between $0$ and $5$ for $1\leq k\leq a$.  Such a number of conservation laws would be appropriate for modeling hydrocarbon combustion, for example.  As a reference, we also consider the case of no conservation laws. For each reaction, we randomly choose the number of reactant species and the number of product species to be one or two (with equal probability), and we randomly select stoichiometric coefficients for each reactant and product species to be one or two (with equal probability). We then randomly select from all sets of such reactants and products that satisfy all the atomic conservation laws. Repeated reactions are excluded in the model generation.}

In order to address the problem of quantifying the degree of catalysis in a network,  we simply deem reactions as catalytic if at least one reactant is also a product.  In order to control the degree of catalysis in the network, we initially exclude randomly generated catalytic reactions from the model. To assess the role of catalysis, we fix the proportion $c$ of the reactions in the network that will be made catalytic and randomly select $\lfloor crn \rfloor$ catalytic reactions. For each catalytic reaction $\ell$, we randomly select one of the reactants $i$ and set $\eta_{i\ell} = \nu_{i\ell}$ to ensure that this species $i$ acts as a catalyst for the reaction. Larger concentrations of a catalytic species thus accelerate both the forward and the reverse rates of the reactions without impacting the stoichiometric balance. We thus generate reaction networks with $\lfloor crn \rfloor$ catalytic reactions, ensuring that the remaining $\lceil (1-c)rn \rceil$ reactions have distinct reactants and products.

\subsection{{Chemical reaction network percolation}}
\label{percolationsec}
Classical percolation theory is based on Erd\H{o}s-R\'{e}nyi graphs, in which each graph of $N$ nodes with $M$ edges is equally likely.\cite{2018_newman} As the connectivity $M/N$ increases for large $N$, this ensemble of networks undergoes a series of percolation transitions. First, as $M/N$ increases above $1/2$,  the expected size of the largest connected component becomes a finite fraction of the network, and a giant connected component is said to emerge. Second, as $M/N$ increases above $\log(N)/2$, the probability of disconnected components in the network almost certainly vanishes. 

For a large number of species $n$, our undriven networks undergo a series of percolation-like transitions with increasing $r$. These transitions differ from those in the classical percolation of Erd\H{o}s-R\'{e}nyi networks, as shown in Fig.~\ref{fig2}, because, as noted above, the addition of each chemical reaction is represented by a collection of links in the network connecting all reactants and products in the reaction. Such a group of fully connected nodes is called a clique, and each clique of $m$ species corresponds to $m(m-1)/2$ links in the network. {In our chemical networks with $a=0$, each reaction has a $1/4$th chance of producing a $2$-species clique, a $2/4$th chance of producing a $3$-species clique, and $1/4$th change of producing a $4$-species clique, leading to an average of $13/4$ links per reaction. The classic results for  Erd\H{o}s-R\'{e}nyi graphs would then imply that the appearance of a giant connected component corresponds to $r>0.5/(13/4) \approx 0.15$, and the onset of full connectivity would (almost certainly) correspond to $r / \log(n) > 0.5/(13/4)\approx 0.15$. For $a>0$, the atomic conservation constraints alter the distribution of cliques introduced with each reaction. Because it is harder to satisfy the constraints when fewer reactants and products are involved, the networks are biased toward more links per reaction. }

\begin{figure*}
\begin{center}
\includegraphics[width=\linewidth]{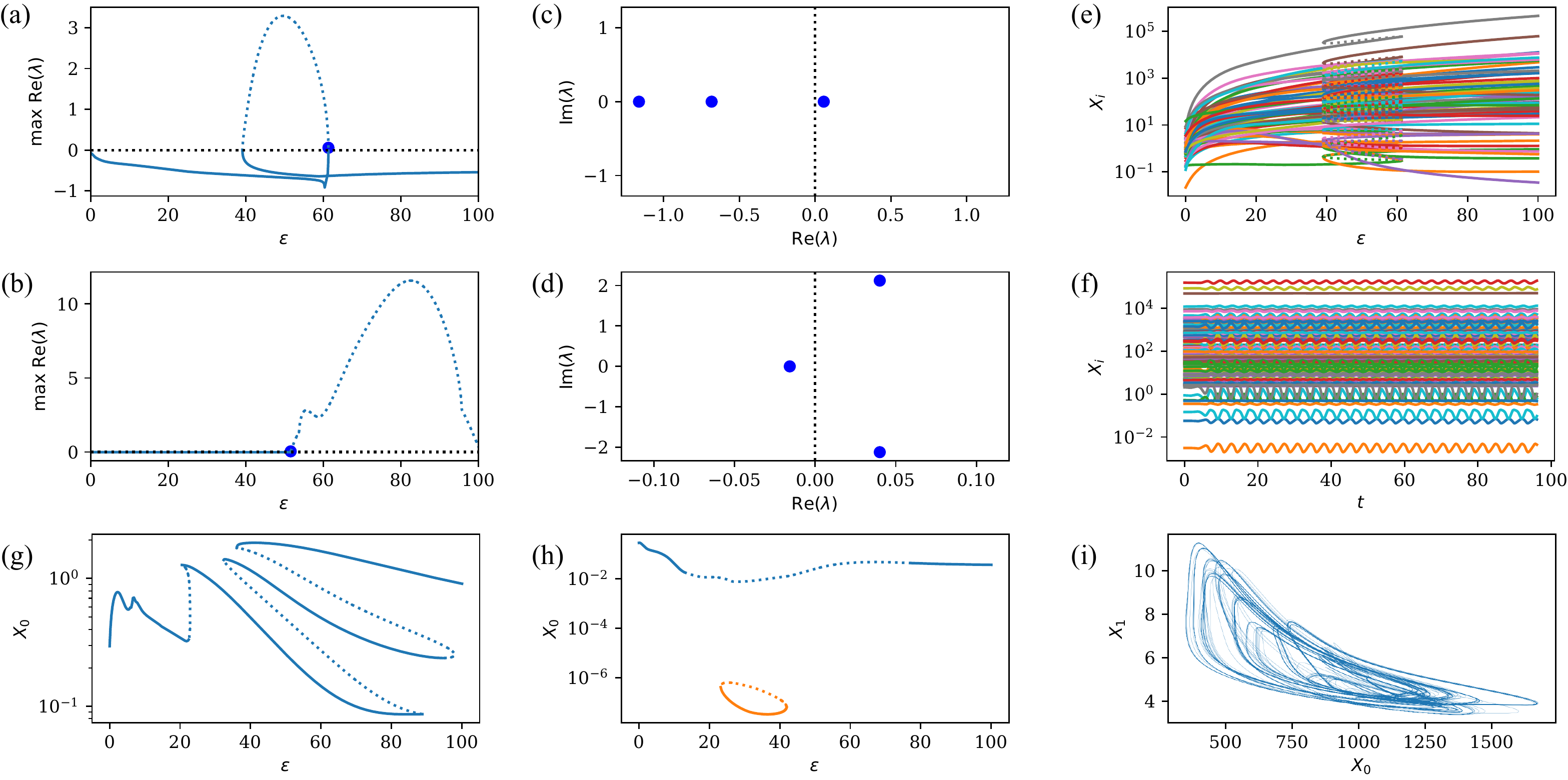}
\end{center}
\caption{Bifurcations resulting from driving in chemical networks. (a) and (b) The maximum real part of the spectrum vs.\ the driving strength $\varepsilon$ for a system undergoing a saddle-node bifurcation (a) and a Hopf bifurcation (b). (c) and (d) Eigenvalues for values of $\epsilon$ [marked by dots in (a) and (b)] just above critical driving $\varepsilon_c$ at which the solutions become unstable for the systems in (a) and (b). (e) Multistable steady state solution concentrations $X_i$ vs.\ driving strength $\varepsilon$ for the system in (a), with stable (unstable) solutions shown as solid (dashed) lines.  (f) Convergence of the concentrations as a function of time to the limit-cycle attractor that emerges from the supercritical Hopf bifurcation in (b). (g) Steady state concentration $X_0$ as a function of driving strength for a different system undergoing a series of snaking bifurcations.  (h) Steady state concentration $X_0$ as a function of driving strength exhibiting a secondary branch of solutions (orange lines). (i) Chaotic attractor observed in a catalytic network with $\varepsilon=75$.  The systems here consist of $n=128$ species with $r/\log(n)=0.5$, $d=0.1$,  $a=3$, and $c=0$ [(a)-(f)] or $c=0.5$ [(g)-(i)]. \label{fig3}}
\end{figure*}

Figures \ref{fig2}(a)--\ref{fig2}(c) shows the average proportion of the network contained in the largest connected component for  Erd\H{o}s-R\'{e}nyi networks and chemical networks with the same number of expected links corresponding to the parameter $r$. For small $r$, the size of the largest connected component decreases with increasing network size, but for $r$ above a critical value of $r=0.075$, a giant connected component appears in each case. As $n\to\infty$, this transition becomes more abrupt and, ultimately, the proportion of the network contained within the largest connected component is zero for $r< 0.075$ and becomes positive for $r> 0.075$. Thus, in terms of the appearance of the giant connected component, chemical networks behave similarly to Erd\H{o}s-R\'{e}nyi networks.

Figures \ref{fig2}(d)--\ref{fig2}(f) shows the proportion of fully connected networks (blue lines) and the proportion of networks with maximal-rank $M_{ij}$ (accounting for the conservation laws) when restricted to the giant connected component (orange lines) as a function of $r/\log(n)$ for Erd\H{o}s-R\'{e}nyi networks and chemical networks. In all cases, the percolation transition from singular to full rank $M_{ij}$ precedes the transition from disconnected to connected, and all transitions sharpen with increasing $n$, leading to sharp transitions in the $n\to\infty$ limit.
For these connectivity transitions, however, the chemical networks differ significantly from the Erd\H{o}s-R\'{e}nyi networks. While the Erd\H{o}s-R\'{e}nyi networks exhibit the transition at the expected $r/\log(n)=0.075$, the chemical networks exhibit the transitions at much larger $r/\log(n)$.
Physically, the delayed crossover for chemical reaction networks is a consequence of correlations in the network structure encoding the chemistry. Since links in the chemical reaction networks are introduced in cliques rather than independently, the connectivity is more localized than for Erd\H{o}s-R\'{e}nyi networks, and additional reactions are therefore required to connect the entire network to the giant connected component.
In real reaction networks, we often expect a sufficiently large number of reactions that the steady state will be unique (neglecting the stoichiometric conservation laws related to the conservation of atomic species here for simplicity). This follows since increasingly rare reactions can always be introduced to the mechanism provided that they satisfy conservation constraints, implying that the actual value of $r$ is very large.
In order to ensure that almost all networks have nonsingular $M_{ij}$ when restricted to the giant connected component (and, consequently, a unique steady state), we assume that $r/\log(n)\geq 0.5$ and $n\geq 64$ in what follows.

\subsection{Emergence of complexity under driving}
\label{complexitysec}
Now we turn to understanding the effects of driving on random reaction networks. To do so, we quasistatically increase the driving amplitude $\varepsilon$ and study the quasistatic response of the undriven steady state (see Appendix \ref{aquasistatic}).  Figure \ref{fig3} shows the impact of driving in three sample networks. The driving alters the spectrum of the Jacobian qualitatively by both shifting the real parts of the eigenvalues and, importantly, allowing eigenvalues to acquire imaginary components.  For sufficiently large driving, the leading eigenvalue may acquire a nonnegative real component at a critical $\varepsilon=\varepsilon_c$, and the steady state will then undergo a bifurcation [Figs.~\ref{fig3}(a)-\ref{fig3}(b)]. The real part of the spectrum may be bounded above by either a single real eigenvalue [Fig.~\ref{fig3}(c)] or be a pair of complex conjugate eigenvalues [Fig.~\ref{fig3}(d)], leading to two qualitatively different bifurcations. Other one-parameter bifurcations, such as the pitchfork bifurcation, are not structurally stable against random perturbations and, hence, are not observed in our random networks.\cite{2013_Kuznetsov}

If a single real eigenvalue crosses the imaginary axis with increasing driving [Fig.~\ref{fig3}(c)],
the system undergoes a saddle-node bifurcation. In this bifurcation, a second, unstable fixed point collides with the stable fixed point $X_i^0(\varepsilon)$ when $\varepsilon=\varepsilon_c$ and both fixed points vanish for $\varepsilon>\varepsilon_c$. After the decay of an irreversible transient, the chemical concentrations stabilize around a different secondary branch of attracting solutions, as shown by the solid lines in Fig.~\ref{fig3}(e).
If the driving were then relaxed back to $\varepsilon<\varepsilon_c$,
the chemical concentrations would not immediately return to the original equilibrium but would remain on this secondary solution branch as long as it continues to exist.
Thus, the appearance of a saddle-node bifurcation can imply the emergence of multistability and hysteresis, enabling chemical memory of the previous driving conditions for the system.

If, instead, two complex-conjugate eigenvalues simultaneously acquire positive real components for sufficiently large driving [Fig.~\ref{fig3}(d)], the system undergoes a Hopf bifurcation. In this bifurcation, a limit cycle coincides with the fixed point at $\varepsilon=\varepsilon_c$, and the fixed point becomes unstable for $\varepsilon>\varepsilon_c$. This bifurcation may be supercritical, meaning that the limit cycle emerges as a stable attractor for $\varepsilon>\varepsilon_c$. In the supercritical case, the concentrations begin to oscillate periodically around the limit cycle for $\varepsilon>\varepsilon_c$ [Fig.~\ref{fig3}(f)], and the system would reversibly return to $X_i^0(\varepsilon)$ if the driving were relaxed to $\varepsilon<\varepsilon_c$. This form of bifurcation thus enables nonstationary dynamics to emerge in response to driving. Alternatively, the Hopf bifurcation may be subcritical, meaning that the limit cycle emerges as an unstable state for $\varepsilon<\varepsilon_c$. As in the saddle-node bifurcation [Fig.~\ref{fig3}(e)], the concentrations will stabilize around a different attractor after the decay of a transient for $\varepsilon>\varepsilon_c$ in the subcritical case, and the system exhibits hysteresis if the driving is decreased again to $\varepsilon < \varepsilon_c$.

As the driving intensity increases beyond an initial bifurcation, additional bifurcations are likely to occur, resulting in yet more complex dynamics. We observe that these behaviors are more common for larger values of the catalytic parameter $c$. For example, solution branches often undergo a snaking series of saddle-node and Hopf bifurcations [blue line in Fig.~\ref{fig3}(g)], resulting in the coexistence of a plethora attracting steady states and limit cycles. Furthermore, secondary solutions branches that are not connected to the original branch may appear [orange line in Fig.~\ref{fig3}(h)]. Finally, the limit cycles emerging from Hopf bifurcations can undergo subsequent period-doubling and torus bifurcations, which can eventually result in chaotic behavior. Although it can be difficult to rigorously confirm chaos in high-dimensional systems, chaotic attractors apparently emerge in our networks [Fig.~\ref{fig3}(i)] and have been previously experimentally confirmed in specific chemical networks~\cite{1987_Argoul}.
\begin{figure*}
\begin{center}
\includegraphics[width=0.65\linewidth]{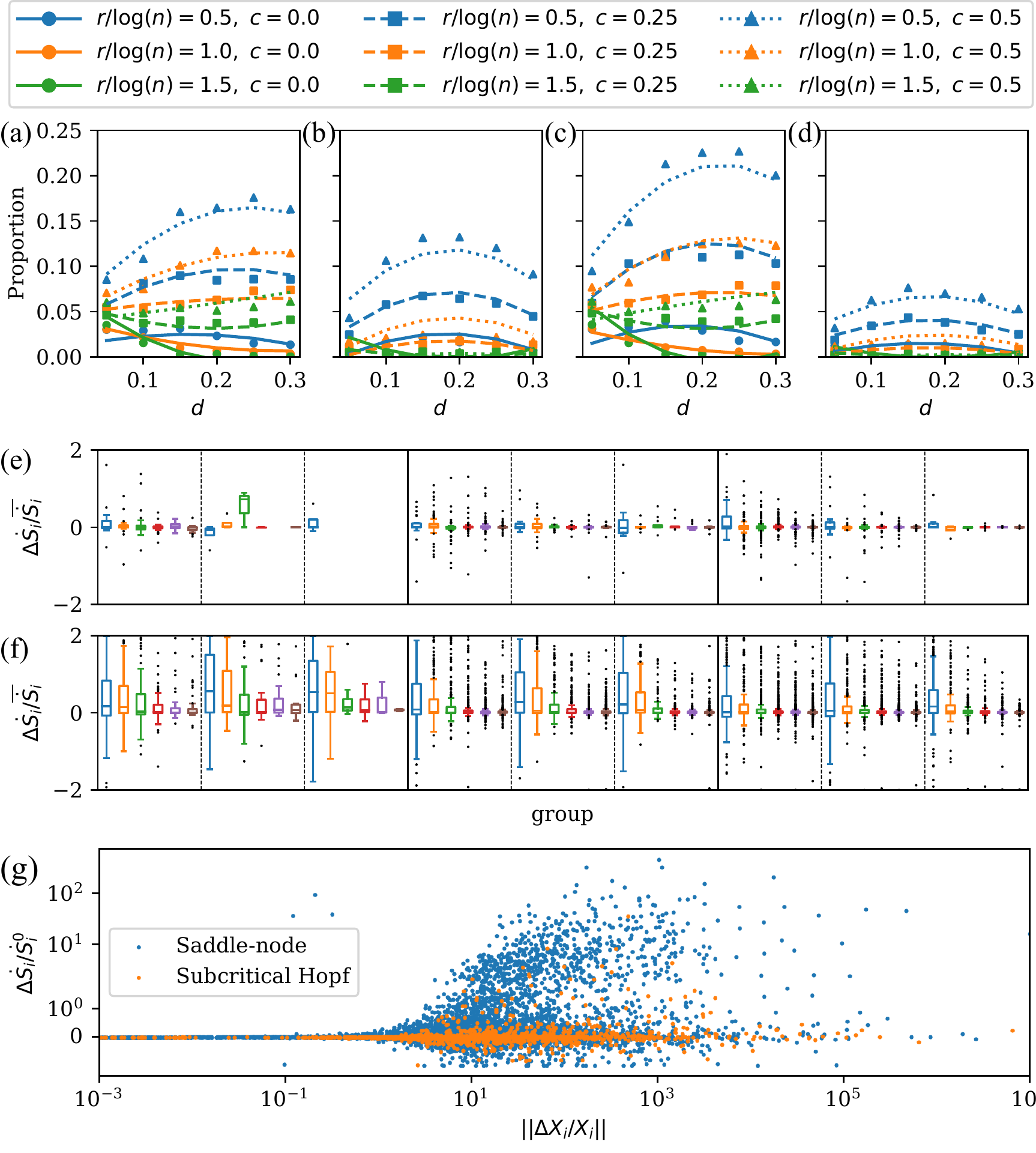}
\end{center}
\caption{Correlations between thermodynamics and complexity. (a)-(d) Proportion of networks exhibiting saddle-node bifurcations (a) and Hopf bifurcations (b) and relaxing to steady states (c) and oscillatory states (d) following a bifurcation, as a function of the scaled number of reactions, proportion of catalytic reactions, and driving proportion. Lines show the result of a nonlinear least-squares fit to a cubic polynomial and each point represents $4096$ random networks for the corresponding parameter combination with $n=128$ species. (e) and (f) Box plots of the relative entropy change for each of the groups of parameters are shown in (a)-(d) for subcritical Hopf bifurcations (e) and saddle-node bifurcations (f).  The groups with differing $c$ and $r/\log(n)$ are divided by solid lines and dashed lines, respectively, and ordered by magnitude, while the groups differing $d$ are distinguished by color.  Cases with no boxes represent groups that exhibited no such bifurcations. (g) Relative change in the entropy production rate vs.\ change in chemical concentrations for the saddle-node and subcritical Hopf bifurcations in (e) and (f), on a symlog scale. \label{fig4}}
\end{figure*}

\subsection{Factors governing the development of complexity}
\label{factorssec}
To understand the impact of the number of reactions, the proportion of catalytic reactions, and the chemical driving on the development of complexity, we consider random networks of $n=128$ species and $a=3$ conservation laws for various parameter values. For each combination of three parameters values for $r/\log(n)$, $c$, and $d$, we numerically continue the undriven steady state in $4096$ random networks starting from $\varepsilon=0$ until either a bifurcation occurs or we reach $\varepsilon=100$. Following a bifurcation, we numerically integrate the dynamics in Eq.~\eqref{kinetics} just above the critical driving amplitude starting from the last stable state of $X_i^0$ until the system relaxes to a new attractor $X_i'$, which may be a steady state or a time-dependent (periodic, quasiperiodic, or chaotic) oscillatory state.

Figures \ref{fig4}(a)-4(d) show the proportions of these networks exhibiting saddle-node and Hopf bifurcations and the proportions of these networks relaxing to steady and oscillatory states following a bifurcation as a function of the parameters. We observe several trends in these data. First, the proportions of systems exhibiting Hopf and saddle-node bifurcations and relaxing toward steady states and oscillatory states are largest for sparse systems and decrease monotonically with $r/\log(n)$. Second, these proportions increase monotonically with the catalytic parameter $c$. This confirms the prevailing hypothesis that catalysis underlies the emergence of complexity in general systems beyond the previously known examples. Third, these proportions show weaker, nonmonotonic dependence on the proportion of driven species $d$, with most systems showing the maximum proportion for intermediate $0.1<d<0.3$.  The proportion of systems exhibiting Hopf bifurcations and relaxing toward steady states exhibit peaks at a slightly larger driving proportion than that of systems exhibiting Hopf bifurcation and relaxing toward oscillatory states. Lastly, saddle-node bifurcations and systems relaxing toward steady states are more common than Hopf bifurcations and systems relaxing toward oscillatory states, and the distribution for individual bifurcation types and relaxation dynamics across parameters all follow similar trends.

To better quantify the thermodynamics of these bifurcations, we consider the entropy density production rate,
\begin{equation}
\label{entropy}
\frac{\dot{S}_{\mathrm{i}}}{V} = R\sum_\ell \left(j^+_\ell - j^-_\ell\right)\log\frac{j^+_\ell}{j^-_\ell},
\end{equation}
which quantifies the extent of detailed balance breaking in driven reaction networks~\cite{2016_rao}. We find that close to equilibrium in undriven chemical networks, the dynamics follow a gradient descent governed by $\dot{S}_\mathrm{i}$ (see Appendix \ref{agradient}), and we expect that the rate of entropy production will play a role in dynamics far from equilibrium as well (we average the entropy production rate over long times to obtain the relevant time-averaged quantity for oscillatory states). To assess this hypothesis, we consider the change in entropy production rate $\Delta \dot{S}_{\mathrm{i}} \equiv \dot{S}'_{\mathrm{i}}-\dot{S}^0_{\mathrm{i}}$ between the time-averaged entropy production rate of the state $X_i'$ after bifurcation, $\dot{S}'_{\mathrm{i}}$, and that of $X_i^0$ just before the bifurcation, $\dot{S}^0_{\mathrm{i}}$. Figures 4(e) and 4(f) show box-plots of $\Delta \dot{S}_{\mathrm{i}} $ relative to the average entropy production rates of the initial and final states $\overline{\dot{S}_{\mathrm{i}}}\equiv \left(\dot{S}'_{\mathrm{i}}+\dot{S}^0_{\mathrm{i}}\right)/2$ for subcritical Hopf bifurcations (top row) and saddle-node bifurcations (bottom row) with each of the parameter combinations in Figs. ~4(a)-4(d). (We exclude supercritical Hopf bifurcations in this analysis since the time-averaged entropy production rate of the stable limit cycle does not differ from the initial steady state significantly.) Since all entropy production rates are positive, this relative change is restricted to the interval $[-2,2]$, which aids in visualizing trends. While many bifurcations produce little change in the entropy production rate, there appears to be a bias toward positive change in entropy production rates, especially in the outliers.

Physically, it is more relevant to consider the change in entropy production rate relative to $\dot{S}_i^0$ rather than $\overline{\dot{S}_{\mathrm{i}}}$, since the former defines the most relevant scale of $S_\mathrm{i}$ prior to the bifurcation. Figure \ref{fig4}(g) shows the change in the entropy production rate relative to the initial entropy production rate as a function of the $L_2$-norm of the change in the time-averaged concentrations of the states, $\Delta X_i$. We see first that many bifurcations result in very small $\lVert \Delta X_i / X_i\rVert$ and almost no change in entropy production. These cases occur when the unstable states that emerge from the bifurcation immediately turn around and become stable through a secondary bifurcation. For bifurcation producing significant changes, on the other hand, it is clear that the relative entropy production rate tends to increase more than it tends to decrease. In fact, there is a statistically significant bias toward increasing entropy production rates following a bifurcation in the accumulated data, with $57\%$ of subcritical Hopf bifurcations and $69\%$ of saddle-node bifurcations exhibiting positive $\Delta \dot{S}_{\mathrm{i}}$ and $p<10^{-10}$ in one-tailed binomial tests for both cases. 

We can understand the bias in the change in the entropy production rate by recognizing that the secondary solution branches the system realizes are expected to be farther separated from equilibrium than the initial state.  The evolution of $\dot{S}_\mathrm{i}$ can be naively modeled as a randomly fluctuating variable that evolves along the continuation of the equilibrium solution. Since $\dot{S}_\mathrm{i}$ is strictly positive, its value typically increases as one travels along the continuation. The secondary solution branches that the system reaches following a bifurcation often appear farther along the continuation than the primary solution branch following a secondary bifurcation. In Figure \ref{fig3}(e), for example, the unstable state that emerges from the primary saddle-node bifurcation undergoes a secondary saddle-node bifurcation farther along the continuation, leading to the secondary branch of stable solutions which is expected to have a larger entropy production rate. These results generalize and provide a potential mechanism for previous observation that emergent attractors in chemical networks are fine-tuned to dissipate energy from a driving apparatus~\cite{2017_Horowitz}.

\section{Discussion}
\label{discussionsec}
Our results demonstrate that design is not required to achieve complex dynamics in chemical networks, but rather that such behavior can emerge naturally under driving. We found that complexity arises through bifurcations most easily in sparsely connected networks, and we confirmed the prevailing hypothesis that catalysis promotes the emergence of complex dynamics. Importantly, we established that, in the absence of driving, relaxation toward equilibrium is governed by equations with a gradient form determined by the entropy production rate, while in the presence of driving, bifurcations leading to complex dynamics exhibit a small but statistically significant bias toward increasing the rate of entropy production. This bias is expected to further increase with the driving intensity, given that secondary solution branches are expected to become increasingly separated from the equilibrium state as the driving strength is increased. Our percolation results also suggest a simple criterion $n_r\gtrsim 0.25n\log(n)$ for assessing if a mechanism contains enough reactions to likely represent a connected chemical reaction network, c.f. Figs. ~2(e) and 2(f).

The study of biological systems such as metabolic networks may benefit from our findings. The increasing tendency for sparse networks to exhibit complex behavior may help explain some of the specific structures observed in biochemical networks, in contrast to the less sparse networks seen in combustion and atmospheric chemical reaction mechanics.  Future studies may thus benefit from focusing on the small $r$ limit of our random chemical reaction networks. 
While stochastic thermodynamics~\cite{2017_Ciliberto,2017_Jarzynski} is likely necessary to realistically describe the dynamics of individual cells, we expect that the entropy production rate may play a role in the development of complexity across scales.
This perspective may give insights into important biological phenomena, such as synchronization phenomena relevant to the circadian rhythm~\cite{2012_Mohawk} and cell division~\cite{2018_Martins} that emerge from coupling between spatial compartmentalization~\cite{2019_Laurent} and emergent oscillations. 
We expect that accounting for spatiotemporal dynamics, such as diffusion and hydrodynamics and noise~\cite{2020_Nicolaou}, will give rise to a plethora of synchronous and pattern-forming phenomena in randomly driven chemical networks.
Although we observed only a small tendency for driving to increase the entropy production rate, this bias has the potential to significantly promote the development of complexity over evolutionary time scales. It is thus tempting to speculate that our results may ultimately contribute to our understanding of the emergence of life, where the need to preserve hereditary information despite a tendency to increase entropy production has been previously implicated in symmetry-breaking processes leading to the formation of a primordial genome~\cite{2017_takeuchi}.

\medskip
\noindent\textbf{Acknowledgements} \par 
This publication was made possible, in part, through the support of grants from the John Templeton Foundation and the National Science Foundation under Grant No. 1856250.
It is also based, in part, upon work supported by the U.S.\ Army Research Laboratory and the U.S.\ Army Research Office under grant number W911NF-14-1-0359. Z.G.N.\ is a Washington Research Foundation Postdoctoral Fellow.
The authors acknowledge the use of the supercomputing facilities managed by the Research Computing Group at the University of Massachusetts Boston and the University of Massachusetts Green High Performance Computing Cluster.

\medskip
\noindent\textbf{Author Contributions} \par 
\noindent\textbf{Zachary G. Nicolaou}: Conceptualization (equal); Formal analysis (lead); Investigation (equal); Methodology (equal); Software (equal); Visualization (equal); Writing -- original draft (lead); Writing -- review \& editing (equal). \textbf{Schuyler B. Nicholson}: Formal analysis (supporting); Investigation (equal); Writing -- review \& editing (equal). \textbf{Adilson E. Motter}: Conceptualization (equal); Investigation (equal); Writing -- review \& editing (equal). \textbf{Jason R. Green}: Conceptualization (equal); Funding acquisition (lead); Investigation (equal); Project administration (equal); Resources (equal); Supervision (equal); Writing -- review \& editing (equal).

\medskip
\noindent\textbf{Data Availability} \par 
The data that support the findings of this study are openly available in GitHub at \url{https://github.com/znicolaou/rmtchem}.

\appendix
\section{Random chemical network generation}
\label{adetailed}
We assume that each species can be assigned a standard Gibbs energy $G_i$, and we take a local detailed balance assumption that the ratio of the forward and reverse reaction rate constants (the equilibrium constant) is given by the change in Gibbs energy, $\kappa^+_\ell / \kappa^-_\ell = \exp\left[{-\sum_i \left(\eta_{i \ell} - \nu_{i \ell}\right)G_i / RT}\right]$.
Motivated by well-characterized gas-phase reactions, such as natural gas combustion and light hydrocarbon pyrolysis\cite{gri,2012_ranzi}, we sample Gibbs energies from a standard normal distribution and forward reaction rates from an exponential distribution. We use a mean $\mu=0$ and a standard deviation $\sigma=1$ for the normal distribution and an exponential scale parameter $\lambda=\sum_i (\eta_{i \ell} - \nu_{i \ell} )G_i / RT$ determined by the change in Gibbs energy, and we fix $RT=1$ for simplicity. Algorithm \ref{reac_alg} shows the pseudocode summarizing our random chemical network generation algorithm. An implementation in Python is also available in our GitHub repository\cite{github}.

\section{Numerical continuation of steady state solutions under chemical driving} 
\label{aquasistatic}
Chemical driving is implemented through nonzero values of the influxes $X_i^{\mathrm{in}}$ and removal rates $R_i$.  We randomly select $\lfloor dn \rfloor$  species to drive (where $d<1$), and set the value of $X_i^{\mathrm{in}}= (1+\varepsilon) X_i^0 R^0$ and $R_i=R^0$, where $\varepsilon$ defines the magnitude of the driving away from equilibrium.  Thus, the initial target value for the chemostat is given by the undriven steady state, $Z_i^0=X_i^0$. We fix $R_0=10^6$ for concreteness, which defines a relatively fast timescale for the driving compared to the chemical kinetics.  For $\varepsilon=0$, the state $X_i^0$ remains a steady state, since the driving vanishes for $X_i=Z_i=(1+\varepsilon) X_i^0$.  Note that with sufficiently many driven species, the chemostat can act as a source and sink for all conserved quantities in the undriven network. Thus, the driving breaks the stoichiometric degeneracy in the steady states and generically fixes $X_i^0$ as the unique steady state for $\varepsilon=0$. While we focus specifically focus on the continuation of $X_i^0$ here, future studies may benefit from considering the continuation of other initial states. Furthermore, since the driving affects the Jacobian only through the negative definite diagonal term $-R_i \delta_{ij}$ in Eq.~\eqref{jacobian}, the state $X_i^0$ is attractive for $\varepsilon=0$ if it is attracting in the absence of the chemostat. 

We then quasistatically increase $\varepsilon$ and use root finding to numerically continue the steady state to $X_i^0(\varepsilon)$ for $\varepsilon>0$. {To do so, we implement a pseudo-arclength continuation method designed specifically for our chemical reaction networks. In this approach, we seek a parameterized family of solutions $(X^0_i(s), \varepsilon(s))$ for both the chemical concentrations and the driving strength in terms of a new parameter $s$ (the pseudo-arclength). The steps taken by the traditional pseudo-arclength continuation are based on an approximate geometric projection for the arclength of the solution curve\cite{1991_Doedel}. However, in our case, the chemical species concentrations can change by orders of magnitude during the continuation, and the solution components can come to quickly dominate the driving parameter in the arclength approximation. }

To overcome this issue, we instead numerically continue using the log quantities $Y_i\equiv \log X_i$. Given a current solution $\bar{Y}_i$ and $\bar{\varepsilon}$, we step the solution forward by solving the extended system of equations
\begin{align}
     X_i^{\mathrm{in}} - R_i X_i + \sum_\ell \left( \eta_{i \ell} - \nu_{i \ell} \right)\left(j^+_\ell - j^-_\ell\right)\rvert_{X_i=\exp(Y_i)} = 0, \\
    \sum_i (Y_i-\bar{Y}_i) dY_i/n + (\varepsilon-\bar{\varepsilon}) d\varepsilon  - ds = 0.
\end{align}
where $ds$ is the pseudo-arclength step size and $(dY_i, d\varepsilon)$ is the ``direction vector''. The key to the pseudo-arclength method is that the Jacobian for the extended system does not become singular at regular solution points, which include both saddle-node and Hopf bifurcation points. This is guaranteed by selecting the direction vector to be in the null space of the matrix $\begin{bmatrix} J_{ij}X_j & X_{i}^{0}R^0\delta^i_{\mathrm{driven}} \end{bmatrix}$, which constitutes the first $n$ rows of the step Jacobian for the extended system, where $\delta^i_{\mathrm{driven}}$ is $1$ for the driven species and $0$ for the undriven species. Since the null space is one-dimensional at regular solution points and the final row of the step Jacobian is determined by the direction vector itself, it follows that the step Jacobian is nonsingular and the continuation can be performed. It then follows that the solution can be continued past saddle-node and Hopf bifurcations efficiently.

Bifurcations are detected by monitoring the eigenvalues of the Jacobian, and supercritical and subcritical Hopf bifurcations are distinguished via the first Lyapunov coefficient\cite{2013_Kuznetsov}. Relevant growth and oscillation timescales are extracted in order to numerically integrate Eq.~\eqref{kinetics} following a bifurcation with an adaptive time-stepping, stiffness switching integrator\cite{github}. 
{The pseudo-arclength continuation works well in most cases but sometimes suffers from finite precision issues. Specifically, the condition number of the step Jacobian can become smaller than the machine precision because of the extreme scale differences in the species concentrations. When this occurs, the direction vector and the stability of the solutions cannot be accurately determined, and the continuation fails. Such failure is also associated with extreme stiffness in numerical integration and disparate timescales in the dynamics. }

\section{Gradient dynamics in the absence of driving}
\label{agradient}
For completeness, we provide a derivation for the gradient form of the dynamics for our undriven chemical reaction networks.  Inserting the undriven steady state $X_i^0$ into Eq.~\eqref{jacobian} and using the detailed balance assumption, the Jacobian for the  steady states in the absence of driving is 
\begin{equation}
J^0_{ij} = \sum_\ell \kappa^+_\ell \left( \eta_{i \ell} - \nu_{i \ell} \right)(\nu_{j \ell}-\eta_{j \ell})\exp\left[\frac{G_j-\sum_k \nu_{k\ell}G_k}{RT}\right].
\end{equation}
The real spectrum of $J^0_{ij}$ follows from the fact that the relaxation to thermodynamic equilibrium is governed by an irreversible gradient descent in undriven chemical networks, as we show here. In particular, let $X_i=X_i^0+\delta_i$ for small $\delta_i$. The linearization of Eq.~\eqref{kinetics} gives $d\delta_i / dt = \sum_j J_{ij}^0 \delta_j$. Taking $u_i\equiv \exp({G_i/2RT})\delta_i$ transforms the linearized dynamics into $d u_i / dt=\sum_j A_{ij}u_j$ with $A_{ij}\equiv \exp\left[{\left(G_i-G_j\right)/2RT}\right]J_{ij}^0$. It is easy to verify that this transformation symmetrizes the system, $A_{ij}=A_{ji}$. Since $J_{ij}^0$ is related to a symmetric matrix by a (nonunitary) similarity transformation, it follows that the eigenvalues of $J_{ij}^0$ are strictly real. Furthermore, it follows from the symmetry of $A_{ij}$ that the evolution can be written in the gradient form $d u_i/dt = dF/du_i$, where $F\equiv \sum_{ij} u_i A_{ij} u_j / 2$. Note that $F$ can be factored into a sum of negative semidefinite terms,
\begin{align}
\label{negative}
F&=-\sum_\ell \kappa_{\ell}^f \left(\sum_i u_i(\eta_{i\ell}-\nu_{i\ell})e^{G_i/2RT}\right)^2\frac{e^{-\sum_k \nu_{k\ell}G_k/RT}}{2},
\end{align}
and so $F\leq 0$. Thus, $A_{ij}$ is negative semidefinite, which confirms that $J_{ij}^0$ has strictly nonpositive eigenvalues. Furthermore, note that
$dF/dt = \sum_i (dF/du_i) \times (du_i/dt) = \sum_i |dF/du_i|^2 \geq 0$.
Thus, the function $F$ is a Lyapunov function for the system, which implies irreversible decay to equilibrium. In fact, the Lyapunov function in Eq.~\eqref{negative} is $1/2RV$ times the linearization of the entropy production rate in Eq.~\eqref{entropy}, which is thought to carry thermodynamic information outside of the linear regime as well\cite{2016_rao}. Outside the linear regime, we find that the entropy production rate and the dynamics can be related via
\begin{align}
\label{sdyn}
\frac{dX_i}{dt} = \frac{X_i}{RV}\frac{d\dot{S}}{dX_i} - \sum_\ell \left(\nu_{i\ell}j^+_\ell-\eta_{i\ell}j^-_\ell\right)\log\frac{j^+_\ell}{j^-_\ell}.
\end{align}
The second term on the right-hand side of Equation \eqref{sdyn} is exactly one-half times the first in the linear regime, leading to the gradient form for the linear dynamics.

\onecolumngrid

\begin{algorithm}[H]
\caption{Generate a random chemical reaction network \label{reac_alg}}
\hspace*{\algorithmicindent} \textbf{Inputs: The \# of species $n$} \\
\hspace*{\algorithmicindent} \textbf{\phantom{Inputs:} The \# of reactions $n_r$} \\
\hspace*{\algorithmicindent} \textbf{\phantom{Inputs:} The \# of conservation laws $n_a$} \\
\hspace*{\algorithmicindent} \textbf{\phantom{Inputs:} The \# of catalytic reactions $n_c$} \\
\hspace*{\algorithmicindent} \textbf{\phantom{Inputs:} The maximum conserved quantity $a_{\mathrm{max}}$} \\
\hspace*{\algorithmicindent} \textbf{\phantom{Inputs:} The maximum iterations ${\mathrm{itmax}}$} \\
\hspace*{\algorithmicindent} \textbf{Output:  Species Gibbs energies $G_i$} \\
\hspace*{\algorithmicindent} \textbf{\phantom{Output:}  Species conserved quantities $A_{ik}$} \\
\hspace*{\algorithmicindent} \textbf{\phantom{Output:}  Reaction rate constants $k_i$} \\
\hspace*{\algorithmicindent} \textbf{\phantom{Output:}  Reaction stoichiometry coefficients $\nu_{il}$ and $\eta_{il}$}
\begin{algorithmic}
    \State $G \gets \textrm{rand\_normal}(\mathrm{mean}=0, \mathrm{var}=1, \textrm{size}=(n))$
    \State $A \gets \textrm{rand\_int}(\mathrm{min}=0,\mathrm{max}=a_{\mathrm{max}},\textrm{size}=(n,n_a))$  
    \State $k \gets \textrm{zeros}(\textrm{size}=(2*n_r))$
    \State $\eta \gets \textrm{zeros}(\textrm{size}=(2*n_r,n))$
    \State $\nu \gets \textrm{zeros}(\textrm{size}=(2*n_r,n))$
    \State $\textrm{previously\_enumerated} \gets [\ ]$ \Comment{Exclude repeated reactions}
    \State $l \gets 0$ \Comment{Reaction index}
    \State $i \gets 0$ \Comment{Iteration number}
    \While{$l<n_r\ \mathrm{and}\ i<\mathrm{itmax}$}
	\State $\mathrm{n\_reac},\mathrm{n\_prod} \gets \mathrm{rand\_int(\mathrm{min}=1,\mathrm{max}=2,\mathrm{size}=(2))}$
	\State $\mathrm{react\_stoi} \gets \mathrm{rand\_int(\mathrm{min}=1,\mathrm{max}=2,\mathrm{size}=(\textrm{n\_reac}))}$
	\State $\mathrm{prod\_stoi} \gets \mathrm{rand\_int(\mathrm{min}=1,\mathrm{max}=2,\mathrm{size}=(\textrm{n\_prod}))}$
	\State $\textrm{reac\_sp\_choices} \gets \textrm{combinations}(\textrm{max}=n, \textrm{num}=\textrm{n\_reac})$ 
	\State $\textrm{prod\_sp\_choices} \gets \textrm{combinations}(\textrm{max}=n, \textrm{num}=\textrm{n\_prod})$ 
	\State $\textrm{reac\_sp} \gets \mathrm{random\_choice}(\textrm{reac\_sp\_choices})$
	\State $\textrm{reac\_atoms} \gets \mathrm{sum}(A[\textrm{reac\_sp}]*\mathrm{reac\_stoi},\mathrm{axis}=0)$
	\State $\textrm{prod\_sp\_choices} \gets \mathrm{cases}\left(\mathrm{prod\_sp}\ \mathrm{in}\ \textrm{prod\_sp\_choices}\ \vert\ \mathrm{sum}(A[\textrm{prod\_sp}]*\mathrm{prod\_stoi},\mathrm{axis}=0) == \mathrm{reac\_atoms}\right)$ 	\Comment{Ensure conservation}
	\State $\textrm{prod\_sp\_choices} \gets \mathrm{cases}\left(\textrm{prod\_sp}\ \mathrm{in}\ \textrm{prod\_sp\_choices}\ \vert\ \textrm{intersect}(\mathrm{prod\_sp}, \mathrm{reac\_sp})==\textrm{None}\right)$ 	\Comment{Exclude catalytic reactions}
	\If{$\mathrm{prod\_sp\_choices}\ \mathrm{is}\ \mathrm{not}\ \mathrm{None}$}
	\State $\textrm{prod\_sp} \gets \mathrm{random\_choice}(\textrm{prod\_sp\_choices})$
	\If{$\mathrm{not}\ \left[\left[\textrm{reac\_sp},\textrm{reac\_atoms}\right],\left[\textrm{prod\_sp},\textrm{prod\_atoms}\right]\right]\ \mathrm{in}\ \textrm{previously\_enumerated}$} \Comment{Add forward and reverse reactions}
		\State $\textrm{previously\_enumerated} \gets \textrm{append}(\textrm{previously\_enumerated}, \left[\left[\textrm{reac\_sp},\textrm{reac\_atoms}\right],\left[\textrm{prod\_sp},\textrm{prod\_atoms}\right]\right])$
		\State $\textrm{previously\_enumerated} \gets \textrm{append}(\textrm{previously\_enumerated}, \left[\left[\textrm{prod\_sp},\textrm{prod\_atoms}\right],\left[\textrm{reac\_sp},\textrm{reac\_atoms}\right]\right])$
		\State $\nu[2*l, \mathrm{reac\_sp}] \gets \mathrm{reac\_stoi}$
		\State $\eta[2*l, \mathrm{prod\_sp}] \gets \mathrm{prod\_stoi}$
		\If{$l<n_c$} \Comment{Make this reaction catalytic}
			\State $\mathrm{cat\_sp}\gets\mathrm{rand\_choice}(\mathrm{union}(\mathrm{reac\_sp},\mathrm{prod\_sp}))$
			\State $\nu[2*l,\mathrm{cat\_sp}] \gets \nu[2*l, \mathrm{cat\_sp}]+1$
			\State $\eta[2*l,\mathrm{cat\_sp}] \gets \eta[2*l, \mathrm{cat\_sp}]+1$
		\EndIf
		\State $\nu[2*l+1, \mathrm{prod\_sp}] \gets \mathrm{prod\_stoi}$
		\State $\eta[2*l+1, \mathrm{reac\_sp}] \gets \mathrm{reac\_stoi}$
		\State $\Delta G \gets \mathrm{sum}(\eta[2*l]*G) - \mathrm{sum}(\nu[2*l]*G)$
		\If{$\Delta G>0$} \Comment{Sample forward reaction rate from exponential}
			\State $k[2*l] \gets \mathrm{rand\_exp}(\Delta G)$
			\State $k[2*l+1] \gets k[2*l] * \mathrm{exp}(-\Delta G)$
		\Else
			\State $k[2*l+1] \gets \mathrm{rand\_exp}(-\Delta G)$
			\State $k[2*l] \gets k[2*l+1] * \mathrm{exp}(\Delta G)$
		\EndIf
		\State $l \gets l+1$
	\EndIf
	\EndIf
	\State $i \gets i+1$

    \EndWhile

    \State \Return $(G,A,k,\nu,\eta)$
\end{algorithmic}
\end{algorithm}

\twocolumngrid

\phantom{}
\clearpage


\end{document}